\documentclass[10pt,revtex4]{article}
\usepackage{amsmath}
\usepackage{amssymb}
\usepackage{amsfonts}
\usepackage{amstext}
\usepackage{amsbsy}
\usepackage[mathscr]{eucal}
\usepackage{graphicx}
\usepackage{color}
\usepackage[all]{xy}
\newcommand{\beq}{\begin{equation}}
\newcommand{\eeq}{\end{equation}}
\newcommand{\beqa}{\begin{eqnarray}}
\newcommand{\eeqa}{\end{eqnarray}}
\newcommand{\ket}[1]{| #1 \rangle}

 \numberwithin{equation}{subsection}

\makeindex
\title{\Large\textbf{Entanglement  dynamics of double Jaynes-Cummings interaction model based on geometric invariants}}

\author{\textit{ Hoshang Heydari and Duy Hoang}\\
        \small\textit{Physics Department, Stockholm university 10691 Stockholm Sweden}\\
\\\small\textit{Email: hoshang@fysik.su.se}}
\begin{document}
\maketitle


\begin{abstract}
We investigate entanglement dynamics and properties of double Jaynes-Cummings model. In particular we study the dynamics of double Jaynes-Cummings model based on  geometric invariants for four qubits. We show that  these geometric invariants provide us with useful information about different types of entanglement during the quantum evolution of the system.
\end{abstract}
\maketitle

\section{Introduction}\label{one}
Quantum entanglement plays an important role in many quantum information and quantum computation tasks. And many well-known quantum systems poses  quantum correlation that need to be characterized and realized for use in different applications. One of these very important quantum systems is Jaynes-Cumming model \cite{JC} in quantum optics that describes the interaction between a single-mode quantized field and a single two-level atom. One advantage of Jaynes-Cumming model is that it is exactly solvable model and we can investigate its entanglement properties analytically. One important application of Jaynes-Cumming model is in quantum computing for realization of quantum registers, namely cavity and circuit quantum electrodynamics.

It is also possible to construct a double Jaynes-Cummings model from two separable Jaynes-Cummings models where atoms $A$ and $B$ interact only with the cavity field $a$ and $b$ respectively. Recently there have been some activities to characterize the entanglement properties of such a double Jaynes-Cummings model \cite{Sainz,Yonac,Han}. 

We treat our model system as a closed system with a unitary evolution and an initial pure product state with the atoms in an entangled two-qubit state and the fields in Fock (vacuum) states. This has the advantage that the evolution can be calculated simply, but we are not considering coupling to the environment. Thus we have restriction on the time e.g.,  the study are limited to evolution over a very short time.

In this paper, we also investigate the entanglement dynamics of double Jaynes-Cummings model based on geometrical invariants. In particular in section \ref{JC} we give an introduction to the double Jaynes-Cummings model and in section \ref{GIN} we review the construction of geometric invariants. Finally in section \ref{DYN} we investigate the entanglement dynamics and properties of the double Jaynes-Cummings model based on geometric invariants and four-tangle which is an important measure of entanglement.

In this paper we will use the following notation for a four-qubit state
\begin{equation}\label{Mstate}
\ket{\Psi}=\sum^{1}_{x_{4}=0}\sum^{1}_{x_{3}=0}\sum^{1}_{x_{2}=0}\sum^{1}_{x_{1}=0}
\Upsilon_{x_{4}x_{3}x_{2} x_{1}} \ket{x_{4}x_{3}x_{2} x_{1}}\in\mathcal{H}_{\mathcal{Q}},
\end{equation}
where
$\mathcal{H}_{\mathcal{Q}}=\mathcal{H}_{\mathcal{Q}_{4}}\otimes
\mathcal{H}_{\mathcal{Q}_{3}}\otimes\mathcal{H}_{\mathcal{Q}_{2}}\otimes\mathcal{H}_{\mathcal{Q}_{1}}=
\mathbb{C}^{2}\otimes\mathbb{C}^{2}\otimes\mathbb{C}^{2}\otimes\mathbb{C}^{2}$ is the Hilbert space of four-qubit state and $\dim \mathcal{H}_{\mathcal{Q}}=16$.

\section{Double Jaynes-Cummings model}\label{JC}
In this section we give a short introduction to double Jaynes-Cummings model following the references  \cite{Yonac,Sainz}. Let $A$ and $B$ be two two-level atoms that interact with
single mode cavity fields $a$ and $b$.  If we suppose that i) each atom-cavity system is isolated; ii) atoms are initially in an entangled state; iii) the cavity are initially in unexcited states. Then the dynamics of this system is described by Jaynes-Cummings Hamiltonian
\begin{eqnarray}
  H_{tot} &=& H_{A}+H_{B}\\\nonumber&=&
  \nu_{A}(a^{\dagger}_{A}a_{A}+\frac{1}{2})+\frac{\omega_{A}}{2}\sigma^{A}_{z}+
  g_{A}(a^{\dagger}_{A}\sigma^{A}_{-}+a_{A}\sigma^{A}_{+})
  \\\nonumber&+&
  \nu_{B}(a^{\dagger}_{B}a_{B}+\frac{1}{2})+\frac{\omega_{B}}{2}\sigma^{B}_{z}+
  g_{B}(a^{\dagger}_{B}\sigma^{B}_{-}+a_{B}\sigma^{B}_{+}),
\end{eqnarray}
where $\omega_{i}$ and $\nu_{i}$, with $i=A,B$ are transition and field frequencies respectively, $g_{i}$ are the coupling constant and $\sigma^{i}_{\pm}$ are spin flip operators. Moreover, the Hamiltonian $H_{A}$ gives  rise to the following unitary operator
\begin{eqnarray}\nonumber
   U_{A}&=&e^{i t H^{0}_{A}}(\cos(\hat{\Omega}_{A} t)
   -i \frac{\sin(\hat{\Omega}_{A} t)}{\hat{\Omega}_{A}}
   (\frac{\Delta_{A}}{2}\sigma^{A}_{z}\\&
   +& g_{A}(a^{\dagger}_{A}\sigma^{A}_{-}+a_{A}\sigma^{A}_{+})),
\end{eqnarray}
where $H^{0}_{A}=\nu_{A}(a^{\dagger}_{A}a_{A}+\frac{1}{2}(\sigma^{A}_{z}+1))$,
\begin{equation}
\hat{\Omega}_{A}=\left(g^{2}_{A}(a^{\dagger}_{A}a_{A}+\frac{1}{2}(\sigma^{A}_{z}+1))+
\frac{\Delta^{2}_{A}}{4}\right)^{\frac{1}{2}},
\end{equation}
and $\Delta_{A}=\omega_{A}-\nu_{A}$ is detuning between the atom and the cavity. Furthermore, the Hamiltonian $H_{B}$ gives a similar unitary operator $U_{B}$ as above one. In this paper we are also consider the following states
$$\ket{\Phi(0)}=\cos\alpha\ket{\uparrow\uparrow}
 +e^{i\beta}\sin\alpha\ket{\downarrow\downarrow},$$$$\ket{\Psi(0)}=\cos\alpha\ket{\uparrow\downarrow}
 +e^{i\beta}\sin\alpha\ket{\uparrow\downarrow},$$
where $0\leq \alpha\leq \frac{\pi}{2}$ and $0\leq \beta\leq \pi$. Next we construct two four-qubit states from these states by apply tensor product of above unitary operators for atom $A$ and $B$  as follows
\begin{eqnarray}\nonumber
  \ket{\Phi^{'}(t)} &=& (U_{A}\otimes U_{B})(\ket{\Phi(0)}\otimes \ket{00} )\\&=&
    \Upsilon_{0000}\ket{\downarrow\downarrow00}+
  \Upsilon_{0011}\ket{\downarrow\downarrow11}\\\nonumber
  &+&
  \Upsilon_{0110}\ket{\downarrow\uparrow10}+
  \Upsilon_{1001}\ket{\uparrow\downarrow01}+
  \Upsilon_{1100}\ket{\uparrow\uparrow00}\\\nonumber&=&
  h_{A}(t)h_{B}(t)e^{i\beta}\sin\alpha\ket{\downarrow\downarrow00}+
  g_{A}(t)g_{B}(t)\cos\alpha\ket{\downarrow\downarrow11}\\\nonumber&+&
 f_{A}(t)g_{B}(t)\cos\alpha \ket{\downarrow\uparrow10}+
g_{A}(t)f_{B}(t)\cos\alpha\ket{\uparrow\downarrow01}\\\nonumber&+&
f_{A}(t)f_{B}(t)\cos\alpha\ket{\uparrow\uparrow00},
\end{eqnarray}
\begin{eqnarray}\nonumber
  \ket{\Psi^{'}(t)} &=& (U_{A}\otimes U_{B})(\ket{\Psi(0)}\otimes \ket{00} )\\\nonumber&=&
    \Upsilon_{0001}\ket{\downarrow\downarrow01}+
  \Upsilon_{0010}\ket{\downarrow\downarrow10}\\
  &+&
  \Upsilon_{0100}\ket{\downarrow\uparrow00}+
  \Upsilon_{1000}\ket{\uparrow\downarrow00}
  \\\nonumber&=&
  h_{A}(t)g_{B}(t)e^{i\beta}\sin\alpha\ket{\downarrow\downarrow01}+
  g_{A}(t)h_{B}(t)\cos\alpha\ket{\downarrow\downarrow10}\\\nonumber&+&
 h_{A}(t)f_{B}(t)e^{i\beta}\sin\alpha \ket{\downarrow\uparrow00}+
  f_{A}(t)h_{B}(t)\cos\alpha\ket{\uparrow\downarrow00}
\end{eqnarray}
where $f_{i}(t)=e^{-i\nu_{i}t}\left(\cos(\Omega_{i} t)-i \frac{\Delta_{i}}{2}\frac{\sin(\Omega_{i} t)}{\Omega_{i}}\right)$, $g_{i}(t)=-ig_{i}\frac{\sin(\Omega_{i} t)}{\Omega_{i}}e^{-i\nu_{i}t}$,
$h_{i}(t)=e^{-i\frac{\Delta_{i}t}{2}}$, with $i=A,B$ and $\Omega_{i}=\left(g^{2}_{i}+\frac{\Delta^{2}}{4}\right)^{\frac{1}{2}}$ is the Rabi frequencies.

\section{Four qubits invariants}\label{GIN}
In this section we will review the construction of four-qubit invariants presented in \cite{Levay}. Using the notation which we have introduced in  section \ref{one}, we define the following four column vectors,
\begin{equation}
\mathcal{A}\equiv\left(
               \begin{array}{c}
                 \Upsilon_{0} \\
                 \Upsilon_{1}\\
                 \Upsilon_{2} \\
                 \Upsilon_{3} \\
               \end{array}
             \right)~~\mathcal{B}\equiv\left(
               \begin{array}{c}
                 \Upsilon_{4} \\
                 \Upsilon_{5}\\
                 \Upsilon_{6} \\
                 \Upsilon_{7} \\
               \end{array}
             \right)~~\mathcal{C}\equiv\left(
               \begin{array}{c}
                 \Upsilon_{8} \\
                 \Upsilon_{9}\\
                 \Upsilon_{10} \\
                 \Upsilon_{11} \\
               \end{array}
             \right)~~\mathcal{D}\equiv\left(
               \begin{array}{c}
                 \Upsilon_{12} \\
                 \Upsilon_{13}\\
                 \Upsilon_{14} \\
                 \Upsilon_{15} \\
               \end{array}
             \right),
\end{equation}
where $\mathcal{A},\mathcal{B},\mathcal{C},\mathcal{D}\in\mathbb{C}^{4}$. Moreover, let
$g:\mathbb{C}^{4}\times \mathbb{C}^{4}\longrightarrow\mathbb{C}$ be a bilinear form such that
\begin{equation}
(\mathcal{A},\mathcal{B})\mapsto g (\mathcal{A},\mathcal{B})
\equiv\mathcal{A}\cdot\mathcal{B}=g_{\alpha\beta}\mathcal{A}^{\alpha}\mathcal{B}^{\beta}
=\mathcal{A}_{\alpha}\mathcal{B}^{\alpha},
\end{equation}
where $\alpha,\beta=0,1,2,3$, $g=\mathcal{J}\otimes \mathcal{J} $ and $\mathcal{J}=\left(
           \begin{array}{cc}
             0 & 1 \\
             -1 & 0 \\
           \end{array}
         \right)$ which satisfy $\mathcal{J}^{2}=-\mathbb{I}$ and $\mathcal{M}\mathcal{J}\mathcal{M}^{T}=\mathcal{J}$ for $\mathcal{M}\in SL(2,\mathbb{C})$. The SLOCC invariants for four-qubit states are then given by
         \begin{equation}
         I_{1}=\frac{1}{2}(\mathcal{A}\cdot\mathcal{D}-\mathcal{B}\cdot\mathcal{C})
         \end{equation}
       \begin{eqnarray}
         I_{2}&=&\frac{1}{6}[(\mathcal{A}\wedge\mathcal{B})
         \cdot(\mathcal{C}\wedge\mathcal{D})+(\mathcal{A}\wedge\mathcal{C})
         \cdot(\mathcal{B}\wedge\mathcal{D})\\\nonumber&-&
         \frac{1}{2}(\mathcal{A}\wedge\mathcal{D})^{2}-
          \frac{1}{2}(\mathcal{B}\wedge\mathcal{C})^{2}],
         \end{eqnarray}
 \begin{equation}
         I_{3}=\frac{1}{2}(\mathfrak{a}\cdot\mathfrak{d}-\mathfrak{b}\cdot\mathfrak{c})
         ~~~~\text{and}~~~~ I_{4}=\epsilon^{\alpha\beta\gamma\delta}
         \mathcal{A}_{\alpha}\mathcal{B}_{\beta}\mathcal{C}_{\gamma}\mathcal{D}_{\delta},
         \end{equation}
         where e.g., $\mathcal{A}\wedge\mathcal{B}=(\mathcal{A}_{\alpha}\mathcal{B}_{\beta}-
         \mathcal{A}_{\beta}\mathcal{B}_{\alpha})$ is a antisymmetric matrix and
         $$(\mathcal{A}\wedge\mathcal{B})
         \cdot(\mathcal{C}\wedge\mathcal{D})=
         (\mathcal{A}_{\alpha}\mathcal{B}_{\beta}-
         \mathcal{A}_{\beta}\mathcal{B}_{\alpha})
         (\mathcal{C}^{\alpha}\mathcal{D}^{\beta}-
         \mathcal{C}^{\beta}\mathcal{D}^{\alpha}).$$ Moreover,
         $$\mathfrak{a}=(a_{\alpha})=
         (-\epsilon_{\alpha\beta\gamma\delta}B^{\beta}C^{\gamma}D^{\delta}),~~\mathfrak{b}=(b_{\beta})=
         (\epsilon_{\alpha\beta\gamma\delta}A^{\alpha}C^{\gamma}D^{\delta}),$$
         $$\mathfrak{c}=(c_{\gamma})=
         (\epsilon_{\alpha\beta\gamma\delta}A^{\alpha}B^{\beta}D^{\delta}),~~\mathfrak{d}=(d_{\delta})=
         (-\epsilon_{\alpha\beta\gamma\delta}A^{\alpha}B^{\beta}C^{\gamma}).$$
         Note that we can rewrite $I_{2}$ in terms of the Plucker coordinates of a Grassmannian variety as follows
  \begin{equation}
         I_{2}=\frac{1}{6}\mathcal{P}^{\mu\nu\alpha\beta}\mathcal{P}_{\mu\nu\alpha\beta},
         \end{equation}
         where $$\mathcal{P}_{\mu\nu\alpha\beta}=\Upsilon_{\mu\alpha}\Upsilon_{\nu\beta}
         -\Upsilon_{\mu\beta}\Upsilon_{\nu\alpha}.$$
For example $I_{2}$ is explicitly given by
         \begin{eqnarray}\nonumber
         I_{1}=&&\frac{1}{2}(\Upsilon_{0}\Upsilon_{15}-\Upsilon_{1}\Upsilon_{14}-\Upsilon_{2}\Upsilon_{13}+
         \Upsilon_{3}\Upsilon_{12}-\Upsilon_{4}\Upsilon_{11}\\&+&\Upsilon_{5}\Upsilon_{10}+\Upsilon_{6}\Upsilon_{9}-
         \Upsilon_{7}\Upsilon_{8}).
         \end{eqnarray}
          A more detail geometrical meaning of these   invariants are discussed in \cite{Levay}.
Let
 $S=(I^{2}_{4}-I^{2}_{2})+4(I^{2}_{2}-I_{1}I_{3})$
and
$T=(I^{2}_{4}-I^{2}_{2})(I^{2}_{1}-I_{2})+(I_{3}-I_{1}I_{2})^{2}.$
Then the four-determinant is defined by
\begin{equation}
D_{4}=\frac{1}{256}(S^{3}-27T^{2}).
\end{equation}
Let $\ket{\widetilde{\Psi}}=\sigma_{y}\otimes \sigma_{y}\otimes\sigma_{y}\otimes\sigma_{y}\ket{\Psi^{*}}$, where
$
\ket{\Psi^{*}}=\sum^{1}_{x_{4},x_{3},x_{2}, x_{1}=0}
\Upsilon^{*}_{x_{4}x_{3}x_{2} x_{1}} \ket{x_{4}x_{3}x_{2} x_{1}}.
$
Then the four tangle is defined by
\begin{eqnarray}
\tau_{4}(\ket{\Psi})&=&|\langle\Psi\ket{\widetilde{\Psi}})|^{2}\\\nonumber&=&|
\sum \Upsilon_{r_{1}\cdots r_{4}}\Upsilon_{s_{1}\cdots s_{4}}
\Upsilon_{u_{1}\cdots u_{4}}\Upsilon_{v_{1}\cdots v_{4}}
\epsilon_{r_{1}s_{1}}\cdots\epsilon_{r_{4}s_{4}}\\\nonumber&&
\epsilon_{u_{1}v_{1}}\cdots\epsilon_{u_{4}v_{4}}|
\\\nonumber&=&4|\Upsilon_{0}\Upsilon_{15}-\Upsilon_{1}\Upsilon_{14}-\Upsilon_{2}\Upsilon_{13}+
         \Upsilon_{3}\Upsilon_{12}-\Upsilon_{4}\Upsilon_{11}\\\nonumber&+&\Upsilon_{5}\Upsilon_{10}+
      \Upsilon_{6}\Upsilon_{9}-
         \Upsilon_{7}\Upsilon_{8}|^{2}.
\end{eqnarray}
 To be able to interpret  our results we consider the entanglement dynamics of the following state $\ket{\Psi_{GHZ4}}=\cos\alpha\ket{0000}+e^{i\beta}\sin\alpha\ket{1111}$. For this state $\tau_{4}(\ket{\Psi_{GHZ4}})=\sin^{2}2\alpha$,
$I_{1}(\ket{\Psi_{GHZ4}})=\frac{1}{2}\cos\alpha \sin \alpha$, $I_{2}(\ket{\Psi_{GHZ4}})=\frac{1}{24}\sin^{2}2\alpha$, $I_{3}(\ket{\Psi_{GHZ4}})=I_{4}(\ket{GHZ_{4}})=0$.
And in particular for GHZ state
$\ket{GHZ_{4}}=\frac{1}{\sqrt{2}}(\ket{0000}+\ket{1111})$ we have $\tau_{4}(\ket{GHZ_{4}})=1$, $I_{1}(\ket{GHZ_{4}})=\frac{1}{4}$, $I_{2}(\ket{GHZ_{4}})=\frac{1}{24}$, $I_{3}(\ket{GHZ_{4}})=I_{4}(\ket{GHZ_{4}})=0$, and $D_{4}(\ket{GHZ_{4}})=0$.

\section{Dynamics of double Jaynes-Cummings model}\label{DYN}
In this section we investigate the dynamical properties of the double Jaynes-Cummings model using the geometric invariants that we have discussed in the pervious section. We start with analytical expression of the first invariant for the state $\ket{\Phi^{'}(t)}$
\begin{eqnarray}
\nonumber
I_{1}(\ket{\Phi^{'}(t)}) &=& \frac{4g_{A} g_{B}i^2 e^{-2 i t (\text{$\nu
   $A}+\text{$\nu $B})}}{\left(\Delta_{A}^2+4
  g_{A}\right) \left(\Delta_{B}^2+4 g_{B}\right)}\cos ^2(\alpha )\\\nonumber&&
   \sin \left(t
   \sqrt{\frac{\Delta_{A}^2}{4}+g_{A}}\right) \sin \left(t
   \sqrt{\frac{\Delta_{B}^2}{4}+g_{B}}\right)\\\nonumber&&
(\sqrt{\Delta_{A}^2+4 g_{A}} \cos \left(t
   \sqrt{\frac{\Delta_{A}^2}{4}+g_{A}}\right)-\Delta_{A} i \sin
   \left(t \sqrt{\frac{\Delta_{A}^2}{4}+g_{A}}\right))\\&&\nonumber
   (\sqrt{\Delta_{B}^2+4 g_{B}} \cos \left(t
   \sqrt{\frac{\Delta_{B}^2}{4}+g_{B}}\right)-\Delta_{B} i \sin
   \left(t \sqrt{\frac{\Delta_{B}^2}{4}+g_{B}}\right)),
\end{eqnarray}
In case $\Delta_{A}=\Delta_{B}=0$ when we have exact resonance we get
\begin{eqnarray}
\nonumber
I_{1}(\ket{\Phi^{'}(t)}) &=& -e^{-2 i t (\text{$\nu
   $A}+\text{$\nu $B})}\cos ^2(\alpha ) \sin \left(t
   \sqrt{g_{A}}\right)  \sin \left(t
   \sqrt{g_{B}}\right)
   \cos \left(t\sqrt{g_{A}}\right)
   \cos \left(t\sqrt{g_{B}}\right)
   \\\nonumber&=&-\frac{1}{4}
\sqrt{g_{A}g_{B}}e^{-2 i t (\text{$\nu
   $A}+\text{$\nu $B})}\cos ^2(\alpha )
   \sin \left(2t
   \sqrt{g_{A}}\right) \sin \left(2t
   \sqrt{g_{B}}\right).
\end{eqnarray}
 \begin{figure}\label{fig1}
\begin{center}
\includegraphics[scale=0.65]{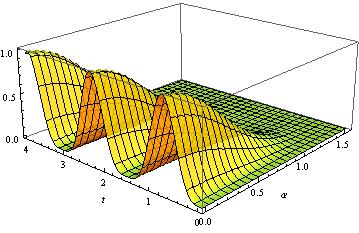}
\end{center}
\caption{Plot of $\tau_{4}((\ket{\Phi^{'}(t)}))$ for the case when we have the same cavity couplings  $g_{A}=g_{B}=1$.}
\end{figure}
Moreover,  when the  relations between cavity couplings is given by  $g_{A}=2g_{B}$ then we get
\begin{eqnarray}
\nonumber
I_{1}(\ket{\Phi^{'}(t)}) &=&\frac{-\sqrt{2}}{4}
  g_{B}e^{-2 i t (\text{$\nu
   $A}+\text{$\nu $B})}\cos ^2(\alpha ) \sin \left(2t
   \sqrt{2g_{B}}\right) \sin \left(2t
   \sqrt{g_{B}}\right).
\end{eqnarray}
Furthermore, when we have same cavity couplings  $g_{A}=g_{B}$, then the above expression is further simplified
\begin{eqnarray}
\nonumber
I_{1}(\ket{\Phi^{'}(t)}) &=&-\frac{1}{4}
  g_{B}e^{-2 i t (\text{$\nu
   $A}+\text{$\nu $B})}\cos ^2(\alpha ) \sin ^2\left(2t
   \sqrt{g_{B}}\right).
\end{eqnarray}
For the above state when $g_{A}=g_{B}=1$ we have plotted the $\tau_{4}((\ket{\Phi^{'}(t)}))=16|I_{1}(\ket{\Phi^{'}(t)})|^{2}$ in Figure 1. Our result also agree with the one given in \cite{Han}.
Next we investigate the entanglement properties $\ket{\Phi^{'}(t)}$ based on the invariant $I_{2}$, that is
\begin{eqnarray}
\nonumber
I_{2}(\ket{\Phi^{'}(t)})   &=& \frac{16g_{A}^2 g_{B}^2 i^4 e^{-4 i t (\text{$\nu
   $A}+\text{$\nu $B})}}{\left(\Delta_{A}^2+4
  g_{A}\right)^2 \left(\Delta_{B}^2+4 g_{B}\right)^2}\\\nonumber&&
   \cos ^4(\alpha ) \sin ^2\left(t
   \sqrt{\frac{\Delta_{A}^2}{4}+g_{A}}\right) \sin ^2\left(t
   \sqrt{\frac{\Delta_{B}^2}{4}+g_{B}}\right)\\\nonumber&&
   (\sqrt{\Delta_{A}^2+4 g_{A}} \cos \left(t
   \sqrt{\frac{\Delta_{A}^2}{4}+g_{A}}\right)-\Delta_{A} i \sin
   \left(t \sqrt{\frac{\Delta_{A}^2}{4}+g_{A}}\right))^2\\&&\nonumber
  (\sqrt{\Delta_{B}^2+4 g_{B}} \cos \left(t
   \sqrt{\frac{\Delta_{B}^2}{4}+g_{B}}\right)-\Delta_{B} i \sin
   \left(t \sqrt{\frac{\Delta_{B}^2}{4}+g_{B}}\right))^2.
\end{eqnarray}
In case when we have exact resonance $\Delta_{A}=\Delta_{B}=0$ we get
\begin{eqnarray}
\nonumber
I_{2}(\ket{\Phi^{'}(t)}) &=& -\frac{1}{16}
\sqrt{g_{A}g_{B}}e^{-4 i t (\text{$\nu
   $A}+\text{$\nu $B})}\cos ^2(\alpha )
   \sin \left(2t
   \sqrt{g_{A}}\right) \sin \left(2t
   \sqrt{g_{B}}\right).
\end{eqnarray}
For this case we can see that $16|I_{2}(\ket{\Phi^{'}(t)})|$ is equal to $\tau_{4}((\ket{\Phi^{'}(t)}))$, see also the plot in Figure 1. First of all we notice that $\tau_{4}((\ket{\Phi^{'}(t)}))$ is depended on $t$, $\alpha$ and coupling constant.  Moreover, from this figure we can observe that the $\tau_{4}((\ket{\Phi^{'}(t)}))$ shows maximal dynamical entanglement if we start from a product state $\ket{\Phi(0)}=\ket{\uparrow\uparrow}\otimes\ket{00}$ for $\alpha=0$ and entanglement vanishes when $\alpha\longrightarrow \frac{\pi}{2}$. This process usually called the entanglement sudden death.

Moreover, we have calculated the invariant  $16I_{4}(\ket{\Phi^{'}(t)})$ which also equal $\tau_{4}((\ket{\Phi^{'}(t)}))$. However, the invariant $I_{3}((\ket{\Phi^{'}(t)}))$ is different from the 4-tangle. Furthermore, for $\Delta_{A}=\Delta_{B}=0$ we have
\begin{eqnarray}
\nonumber
I_{3}(\ket{\Phi^{'}(t)}) &=& -\frac{1}{64}
(g_{A}g_{B})^{\frac{3}{2}}e^{-6 i t (\text{$\nu
   $A}+\text{$\nu $B})}\cos ^6(\alpha )
   \sin^{3} \left(2t
   \sqrt{g_{A}}\right) \sin^{3} \left(2t
   \sqrt{g_{B}}\right).
\end{eqnarray}
The  entanglement of the state also vanishes when $\alpha\longrightarrow \frac{\pi}{2}$ and oscillates with time $t$.
And for a general case of the state
\begin{eqnarray}\nonumber
  \ket{\Phi^{''}(t)} &=&
    \Upsilon_{0000}\ket{\downarrow\downarrow00}+
  \Upsilon_{0011}\ket{\downarrow\downarrow11}+\Upsilon_{0110}\ket{\downarrow\uparrow10}\\\nonumber
  &+&
  \Upsilon_{1001}\ket{\uparrow\downarrow01}+
  \Upsilon_{1100}\ket{\uparrow\uparrow00},
\end{eqnarray}
we get
$I_{1}(\ket{\Phi^{''}(t)})=\frac{1}{2}(\Upsilon_{1001}\Upsilon_{0110}+\Upsilon_{1100}\Upsilon_{0011})$
, $I_{2}(\ket{\Phi^{''}(t)})=\frac{1}{6}(\Upsilon^{2}_{1001}\Upsilon^{2}_{0110}
+4\Upsilon_{1100}\Upsilon_{1001}\Upsilon_{0110}\Upsilon_{0011}+\Upsilon^{2}_{1100}
\Upsilon^{2}_{0011})$
,\begin{eqnarray}I_{2}(\ket{\Phi^{''}(t)})&=&\frac{1}{2}\Upsilon_{1100}\Upsilon_{1001}
\Upsilon_{0110}\Upsilon_{0011}(\Upsilon_{1001}\Upsilon_{0110}+\Upsilon_{1100}\Upsilon_{0011}),\end{eqnarray}
 $I_{4}(\ket{\Phi^{''}(t)})=\Upsilon_{1100}\Upsilon_{1001}\Upsilon_{0110}\Upsilon_{0011}$, and
 $D_{4}(\ket{\Phi^{''}(t)})=0$.
 We also have checked the entanglement properties of $ \ket{\Psi^{'}(t)}$ but all invariants vanish for this state. And even  if we consider a  general state
 \begin{eqnarray}\nonumber
  \ket{\Psi^{''}(t)} &=&
    \Upsilon_{0001}\ket{\downarrow\downarrow01}+
  \Upsilon_{0010}\ket{\downarrow\downarrow10}+  \Upsilon_{0100}\ket{\downarrow\uparrow00}+
  \Upsilon_{1000}\ket{\uparrow\downarrow00},
\end{eqnarray}
then all invariants vanish. Thus all four-qubits invariants that we have discussed in this paper cannot detect any entanglement in $ \ket{\Psi^{'}(t)}$ or $ \ket{\Psi^{''}(t)}$.

In summary we have considered the double Jaynes-Cumming model of interacting atom and quantize field. We have used the geometric invariants to detect entanglement dynamics of this model. We have shown that these geometric invariants revile useful information about entanglement properties of the state $ \ket{\Phi^{'}(t)}$. We also have shown that the state $ \ket{\Phi^{'}(t)}$ obtains  maximal dynamical entanglement if we start from a product state and entanglement of the state  vanishes when $\alpha$ goes to $\pi/2$.  Moreover, we also have found that for the state $ \ket{\Psi^{'}(t)}$ all geometrical invariants including the four-tangle vanish.

\end{document}